\documentclass[aps,prd,twocolumn,superscriptaddress,groupedaddress]{revtex4}  
\usepackage{graphicx}  % needed for figures
\usepackage{dcolumn}   % needed for some tables
\usepackage{bm}        % for math
\usepackage{amssymb}   % for math

\newcommand{\be}{\begin{equation}}
\newcommand{\ee}{\end{equation}}
\newcommand{\bea}{\begin{eqnarray}}
\newcommand{\eea}{\end{eqnarray}}
\newcommand{\bg}{\begin{gather}}
\newcommand{\eg}{\end{gather}}
\newcommand{\bseq}{\begin{subequations}}
\newcommand{\eseq}{\end{subequations}}

\renewcommand{\ln}{\mathop{\rm ln}\nolimits}

\def\ln{{\mathrm{ln}}}

\def\P{{\mathcal{P}}}

\def\P{\mathcal{P}}

\def\p{{\bf p }}

\newcommand{\seq}{\begin{subequations}}
\newcommand{\sen}{\end{subequations}}
\newcommand{\eq}{\begin{eqnarray}}
\newcommand{\en}{\end{eqnarray}}

\def\shiftdown#1{#1\llap{\lower.04ex\hbox{#1}}}

\begin{document}

\title{Photoproduction of axion-like particles in the NA64 experiment}
\author{R.~R.~Dusaev\footnote{{\bf e-mail}: 
renat.dusaev@cern.ch}} 
\affiliation{Tomsk State University,
634050 Tomsk, Russia}
\author{D.~V.~Kirpichnikov\footnote{{\bf e-mail}: kirpich@ms2.inr.ac.ru}}
\affiliation{Institute for Nuclear Research of the Russian Academy 
of Sciences, 117312 Moscow, Russia} 
\author{M.~M.~Kirsanov\footnote{{\bf e-mail}:
mikhail.kirsanov@cern.ch }}
\affiliation{Institute for Nuclear Research of the Russian Academy 
of Sciences, 117312 Moscow, Russia}
                                                        
\date{\today}

\begin{abstract}
Axion-like particles $a$ (ALPs) that couple to the  Standard Model (SM) 
gauge fields could be observed in the high-energy photon
scattering $\gamma N\to N a$ off nuclei followed by the 
$a\to \gamma\gamma$ decay. In the present paper we describe the calculation of the ALP 
production cross-section and the properties of this production. The cross section formulas are implemented
in the program for the simulation of events in the NA64 experiment, the active electron beam dump facility at the CERN SPS.
We study the prospects of the NA64 experiment to search for ALP
in the $10\, \mbox{MeV} \lesssim m_a\lesssim 100$ MeV mass
range for the statistics corresponding to up to $5\times 10^{12}$ electrons on target (EOT).
\end{abstract}

\maketitle

\section{Introduction}

Axion-like particles  interacting with gauge bosons of the 
Standard Model (SM) arise naturally in various well motivated SM extensions 
such as string theory~\cite{Arvanitaki:2009fg,Svrcek:2006yi,Visinelli:2018utg} and 
supersymmetry~\cite{Nelson:1993nf,Bagger:1994hh}.  Being a 
pseudo Nambu-Goldstone boson of spontaneously broken global
Peccei-Quinn symmetry~\cite{Peccei:1977hh}, ALP originally addressed the 
strong-CP problem~\cite{Peccei:1977hh,Weinberg:1977ma,Wilczek:1977pj}. More 
recently the interest to a new light and weakly coupled pseudo-scalar 
particle has been stimulated due to its relevance to the well motivated Dark Matter
(DM) models~\cite{Boehm:2003hm,Dolan:2014ska,Hochberg:2018rjs}.

\begin{figure*}[t]
\begin{center}
\includegraphics[width=0.8\textwidth]{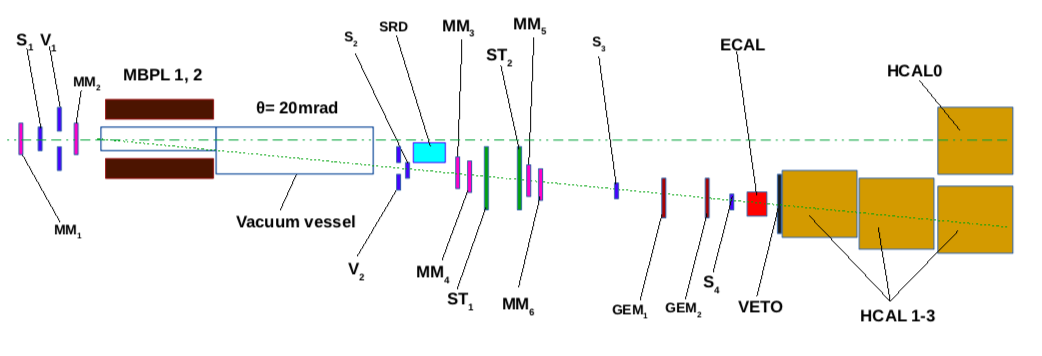}
\caption {The NA64 configuration used for the search for ALP decays $a\to\gamma\gamma$. 
\label{figInvDesgnFull}}
\end{center}
\end{figure*}

The aim of the present work is to study the ALP production in the electron fixed target experiment NA64 at the CERN SPS through the Primakoff 
reaction $\gamma N \to N a$.
The NA64 (Fig.~\ref{figInvDesgnFull}) is an active beam dump facility with a significant potential to probe various scenarios 
beyond the Standard Model (BSM). The well-motivated dark sector
of particle physics has been already constrained by NA64 using the 
missing energy
signatures~\cite{Gninenko:2016kpg,Banerjee:2016tad,Gninenko:2017yus,Gninenko:2018ter,Banerjee:2019invis}.
%Moreover, the new experimental bounds on the $^8$Be$^*$ 
%anomaly~\cite{Krasznahorkay:2015iga} 
%have been derived  from the absence   
%of a new X17 boson decay into electron-positron pairs ~\cite{Banerjee:2018vgk,Banerjee:2019hmi}.
% The authors of Ref.~\cite{Ellwanger:2016wfe} suggested that the %light pseudo-scalar particle $a$
%can be a viable candidate for the explanation of the $^8$Be$^*$ %anomaly. This particle couples predominantly to electrons, 
%such that $\mbox{Br}(a\to e^+e^-)\simeq 1$. The relevant pseudo-%scalar can originate from the UV 
%completed models with vector-like fermions and Higgs extended %sector~\cite{Batell:2016ove,Chen:2015vqy}.
%Probing general scenarios with hidden pseudo-scalar at NA64 is %discussed elsewhere~\cite{KirpPseudoScal}. 
%However, in the present paper we discuss the NA64 expected %sensitivity to 
%pseudo-scalar particles in the simplified benchmark model
%with a dominant coupling to photons and 
%$\mbox{Br}(a\to \gamma \gamma)\simeq 1$.

Probing pseudo-scalar particles in the MeV-GeV mass range by the beam-dump 
facilities is becoming a hot topic for the experimental study. 
For instance, such planned
 experiments as 
FASER~\cite{Ariga:2018uku}, MATHUSLA~\cite{Chou:2016lxi},
SHIP~\cite{Anelli:2015pba}, CodexB~\cite{Gligorov:2017nwh}, 
SeaQuest~\cite{Berlin:2018pwi} and LDMX~\cite{Akesson:2018vlm}
will be able to probe long-lived ALP~\cite{Beacham:2019nyx,Alemany:2019vsk}
due to large distances between the ALP production vertex and detector.
In these experiments
ALP propagates typically along a distance of $10 - 100$~m before its decay.
This implies that the above-mentioned experimental facilities are 
sensitive to relatively small couplings in the range 
$10^{-7} \mbox{ GeV}^{-1} \lesssim  g_{a \gamma \gamma} \lesssim 10^{-4} \mbox{ GeV}^{-1}$. 
On the other hand, the typical decay length in the NA64 experiment is several meters depending on the NA64 geometry configuration.
Therefore, due to this shorter length in this experiment it is possible to search for decays of ALP with
$g_{a \gamma \gamma} \gtrsim 10^{-4}$~GeV for sub-GeV $m_a$. In addition, it is possible to search for long-lived ALP
in the missing energy signatures.

The paper is organised as follows. In Sec.~\ref{PropSect} we discuss the properties of 
ALP. In Sec.~\ref{CSSect} we review the ALP production cross-section 
in the Primakoff reaction. In Sec.~\ref{MCSect} we discuss the Monte-Carlo (MC) simulation 
of the ALP production in the NA64 experiment. In Sec.~\ref{Strategy}
we describe the ALP search strategy. In Sec.~\ref{ExpectedlimitsNA64} we   estimate the expected sensitivity  of NA64
facility to the ALP for the statistics up to $5\times 10^{12}$ electrons on target. We conclude in Sec.~\ref{Concl}.

\section{The ALP properties
\label{PropSect}}

We consider the simplified setup \cite{Dobrich:2015jyk} of ALP coupling predominantly to photons:  
\begin{equation}
\mathcal{L}_{int} \supset - \frac{1}{4} g_{a \gamma \gamma} a F_{\mu \nu} \tilde{F}^{\mu \nu} + 
\frac{1}{2}(\partial_\mu a)^2-\frac{1}{2} m_a^2 a^2,
\label{ALPlagr1}
\end{equation}
where $F_{\mu \nu}$ denotes the strength of the photon field, and the dual tensor is defined by
$\tilde{F}_{\mu \nu}  = \frac{1}{2} \epsilon_{\mu \nu \lambda \rho} F^{\lambda \rho}$.
We assume throughout the paper that the effective coupling, $g_{a \gamma \gamma}$, and the ALP mass, $m_a$, 
are independent. The pseudoscalar boson coupled to photons (\ref{ALPlagr1}) has the following decay width 
\begin{equation}
\Gamma_{a\rightarrow \gamma \gamma} = \frac{g_{a \gamma \gamma}^2 m_a^3}{64 \pi}.
 \label{Width}
\end{equation}  
The decay length of ALP is given by
\begin{equation}
l_a \simeq 4 \mbox{m}\,  \frac{E_a}{10^2\, \mbox{GeV}}\left(\frac{g_{a \gamma \gamma}}{10^{-4} \, \mbox{GeV}^{-1}}\right)^{-2}\!\!
  \left(\frac{m_a}{10^2 \, \mbox{MeV}}\right)^{-4},
 \label{DecLength1}
\end{equation}
where $E_a$ is the ALP energy.
The minimal decay length to which the NA64 facility is sensitive is of the order of the target thickness (0.5m).
Therefore, from Eq.~(\ref{DecLength1}) one can conclude that NA64 with
a most used beam energy of 100 GeV is sensitive to the values of ALP coupling to photons of the order of
$g_{a \gamma \gamma}\gtrsim 10^{-4}\, \mbox{GeV}^{-1}$.
 
\begin{figure*}[t]
\begin{center}
\includegraphics[width=0.496\textwidth]{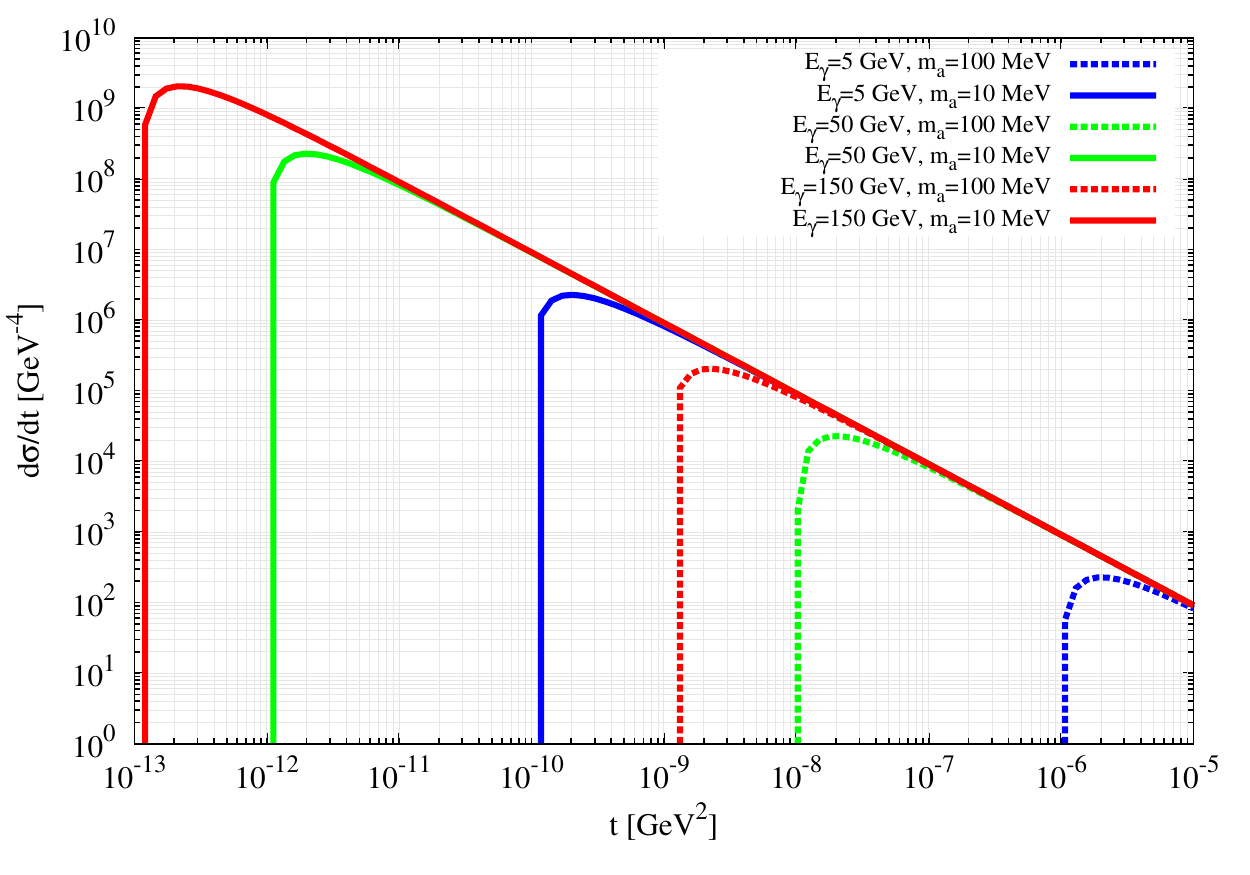}
\includegraphics[width=0.496\textwidth]{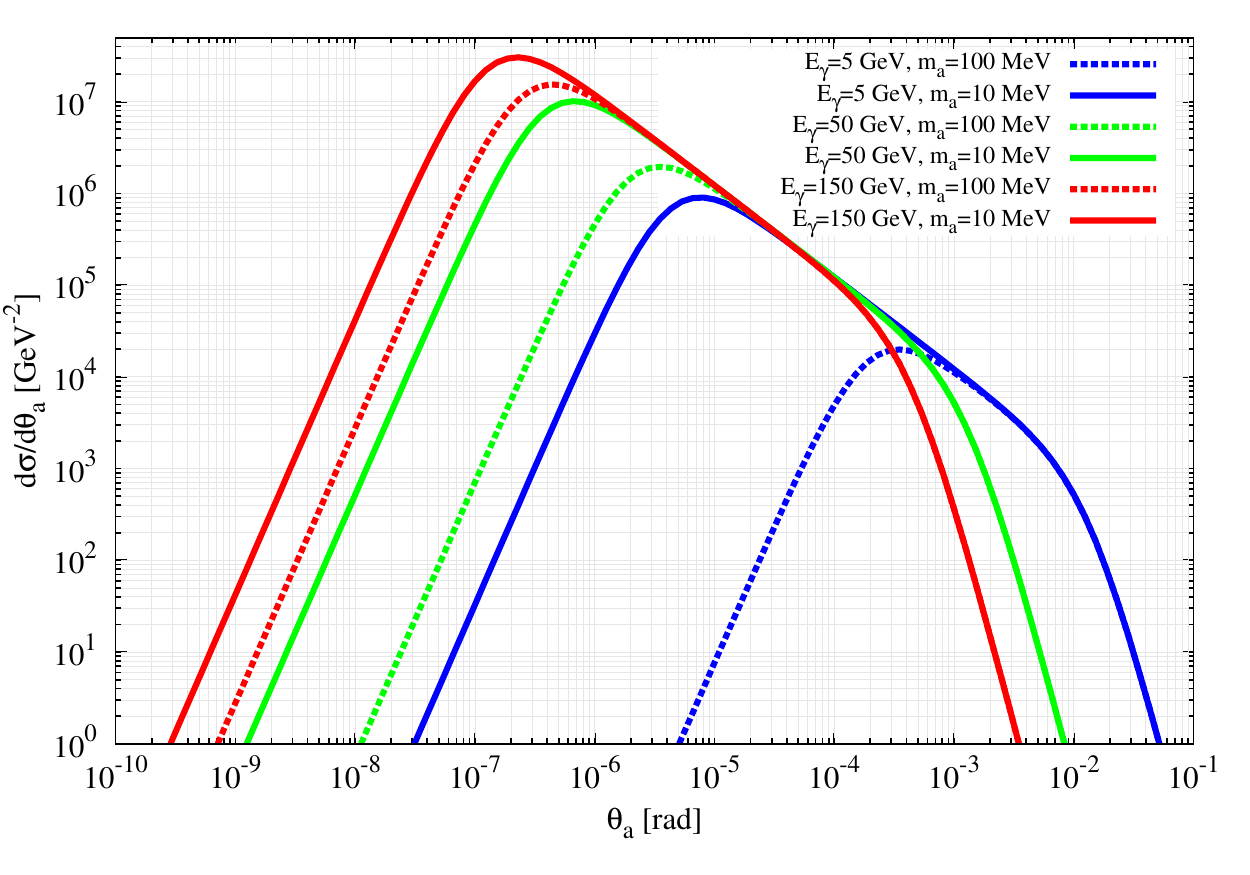}
\caption {Left: differential cross-section versus momentum tranfer squared.
Right: differential cross-section versus angle of ALP emission. All 
cross-sections are calculated for lead target and $g_{a\gamma\gamma}=1\, \mbox{GeV}^{-1}$.}
\label{dsdtANDdsdtheta}
\end{center}
\end{figure*}  

\begin{figure}[t]
\begin{center}
\includegraphics[width=0.5\textwidth]{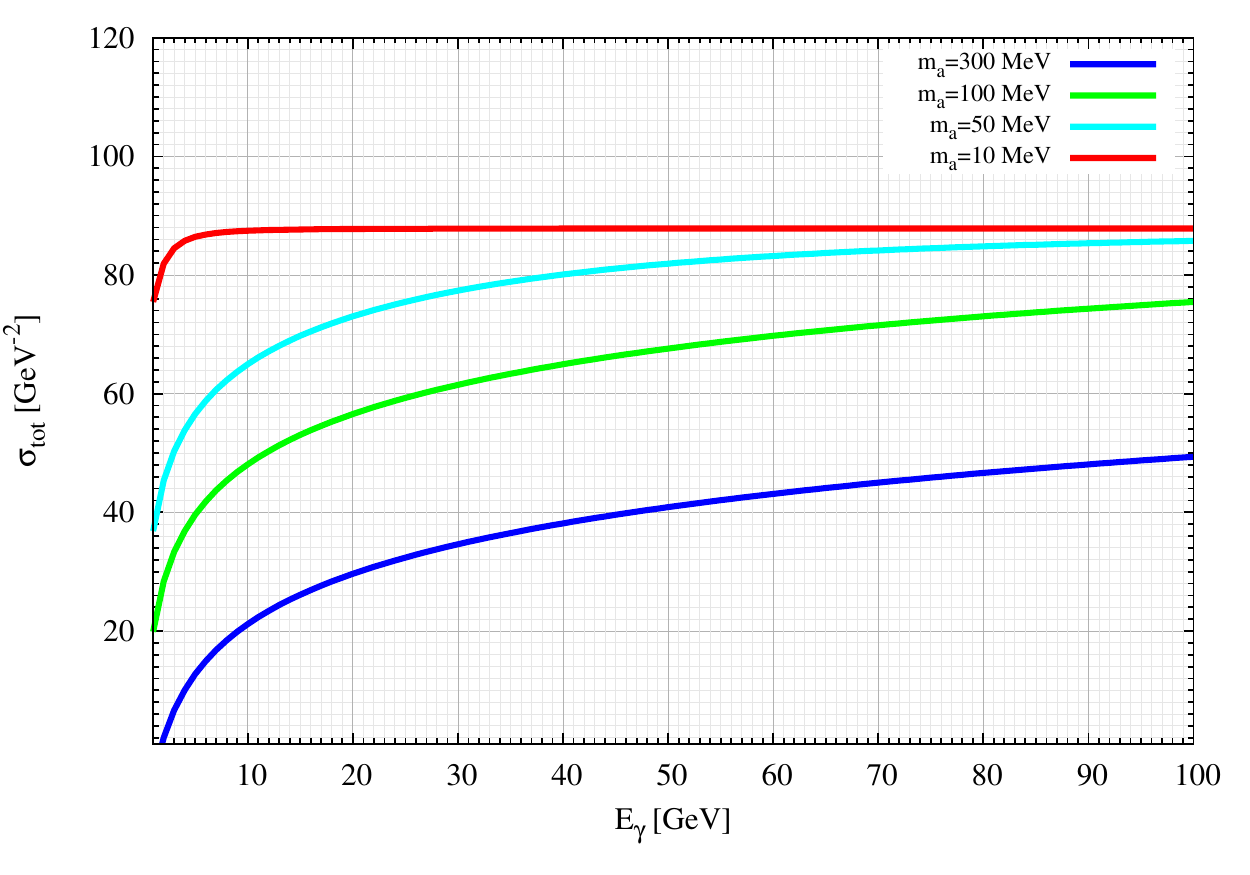}
\caption {Total cross-section versus incident photon energy for lead target and $g_{a\gamma\gamma}=1\, \mbox{GeV}^{-1}$.}
\label{stotVsEg}
\end{center}
\end{figure}

\section{Cross-section
\label{CSSect}}

We first calculate the cross-section of axions produced in the Primakoff
process  $\gamma N\rightarrow a N$.
%The incident photons 
%in this reaction originate from the bremsstrahlung radiation of a primary 
%electron beam degraded in the target. The spectra of these incident photons  
%were simulated in {\tt GEANT4}~\cite{Agostinelli:2002hh} for the corresponding
%design of NA64 facility.
The ALP production amplitude is given by
\begin{equation}
\mathcal{M}= g_{a \gamma \gamma} e  F(q^2)\,
\epsilon_{\mu \nu \lambda \rho}\,\epsilon_{i}^\mu(p)\, p^\lambda\, q^\rho (\P_i+\P_f)^\nu \, \frac{1}{q^2},
\label{ALPprodAMPL1}
\end{equation}
where $p,\P_i,\P_f$ and $k$ are the four-momenta of the incident photon, initial 
nucleus, final state nucleus and the axion respectively, $e$ is the electron charge.
The internal photon momentum is defined by $q=\P_i - \P_f$.
In Eq.(\ref{ALPprodAMPL1}) we suppose that the nucleus has spin zero, thus corresponding nuclear-photon
vertex is given by~\cite{Liu:2017htz,Liu:2016mqv,Beranek:2013nqa,Beranek:2013yqa}
  $$ i e F(q^2) (\P_i +\P_f).$$
The form-factor $F(q^2)$ depends upon 
the value of momentum transfer $q^2=-t$ and
 describes the elastic photon scattering~\cite{Bjorken:2009mm}
\begin{equation}
F(t) \approx Z \left( \frac{a^2t }{1+a^2t} \right) 
\left(\frac{1}{1+t/d}\right),
\label{elFF1}
\end{equation}
where $a=111 Z^{-1/3}/m_e$ and $d=0.164\, \mbox{GeV}^2 A^{-2/3}$,
here $m_e$ is the mass of electron and $A$ is the atomic weight.
The inelastic  form-factor proportional to $\sqrt{Z}$ is small 
as compared to (\ref{elFF1}) for the high-$Z$ target material and
thus yields a subdominant contribution to the ALP production that we neglect.
The differential cross-section of the elastic processes $\gamma N\to N a$ in the lab 
frame is given by
\begin{equation}
d \sigma = \frac{1}{2^5 \pi} \frac{1}{E^2_\gamma M_N}  
\overline{\left| \mathcal{M}\right|^2}\, d E_a, 
\label{DsDEa1}
\end{equation}
where $E_{\gamma}$ is the incoming photon energy and $M_N$ is the mass of nucleus.
The amplitude squared (see, e.~g.~Eq.~(\ref{ALPprodAMPL1}) for details)
averaged over the initial photon polarizations is given by
$$
 \overline{\left| \mathcal{M}\right|^2} = \frac{1}{2} \sum_{pol.} \left| \mathcal{M}\right|^2 = 
g_{a \gamma \gamma}^2 e^2  F^2(q) M_N^2 \,  \frac{1}{2 t^2}\times
$$
\begin{equation}
\left[ (4E_a^2 t -(m_a^2+t)^2) - \frac{2 E_a t (m_a^2-t)}{M_N} -\frac{m_a^2 t^2}{M_N^2} \right]
\label{M2viaEa}
\end{equation}
here we use the {\tt FeynCalc} package~\cite{Shtabovenko:2016sxi} 
of {\tt Wolfram Mathematica}~\cite{Mathematica} that carries out a 
summation in $\overline{\left| \mathcal{M}\right|^2}$  over dump indices. 
The resulting amplitude squared is given by
$$
  \overline{\left| \mathcal{M}\right|^2} \simeq
g_{a \gamma \gamma}^2 e^2  F^2(t) M_N^2 \times  \frac{1}{2 t^2}
 (4E_a^2 t -m_a^4), 
$$
where we suppose that $m_a \gg t$ and neglect the third and fourth terms
of Eq.~(\ref{M2viaEa}) since the target nuclei are 
rather heavy, $M_N\gg m_a$ and $M_N\gg \sqrt{t}$. The angle $\theta_a$ between the incoming photon 
and ALP can be derived from the momentum conservation law.
The  latter implies the following expression
\begin{equation}
\cos \theta_a = \frac{1}{2 |\p_a| E_\gamma} \cdot (2E_a (E_\gamma +M_N) -2E_\gamma M_N -m_a^2 ).
\label{CosThetaA}
\end{equation}
For the NA64 experiment we are mainly interested in high energy 
photons produced by $100$ GeV electrons in the lead target.
This corresponds to small momentum transfers and to small angles of ALP emission. In particular,  
we consider the limit when $m_a \ll E_a$ and $\theta_a \ll 1$, 
then Eq.~(\ref{CosThetaA}) implies that the photon energy can 
be expressed as 
\begin{equation}
E_\gamma \approx E_a +\frac{E_a^2 \theta_a^2}{2 M_N}+ \frac{m_a^4}{8 M_N E_a^2}. 
\end{equation}
In this approach the ALP energy can be rewritten as follows
\begin{equation}
E_a \approx E_\gamma - \frac{E_\gamma^2 \theta_a^2}{2 M_N}- \frac{m_a^4}{8 M_N E_\gamma^2}. 
\label{EaApprox}
\end{equation}
We note that one should not neglect the second term in 
Eq.~(\ref{EaApprox}) which is naively associated with a typical angle of ALP emission. 
In particular, from Eq.~(\ref{EaApprox}) follows that the momentum transfer squared can be approximated as
\begin{equation}
t=-q^2=2M_N (E_\gamma - E_a) \approx 
 E_\gamma^2 \theta_a^2 +\frac{1}{4} \frac{m_a^4}{E_\gamma^2}.
 \label{t1def} 
\end{equation}
It is worth mentioning, however, that the realistic typical angle of ALP production depends also on the
properties of atomic form-factors (see, e.~g.~Eq.~(\ref{typicalAngle}) below for details).
Finally, one can obtain the momentum transfer distribution from Eqs.~(\ref{DsDEa1}) and~(\ref{t1def})
\begin{equation}
\frac{d \sigma}{d t} =  \frac{1}{2^3  }\cdot g_{a \gamma \gamma}^2 \alpha
F^2(t)\cdot \frac{1}{t^2 }\left(t-t_{min}\right) 
\label{dsdt1}
\end{equation}
where $t_{min}=m_a^4/(4E_\gamma^2)$. The differential cross-section $d \sigma /dt$ has a peak at
\begin{equation}
t^*= 2t_{min}+1/a^2,
\end{equation}
which is associated with typical momentum transfers. In the left panel of
Fig.~\ref{dsdtANDdsdtheta} we show $d\sigma/dt$ as a function of $t$ for various
masses $m_a$ and typical energies of incoming photons $E_\gamma$.
We note that the maximum allowed value of momentum transfer is given by
$$q_{max}=\sqrt{t_{max}}=\sqrt{2M_N(E_\gamma-m_a)}.$$
For the typical threshold energy of interest $E_\gamma > 50$ GeV we have $q_{max}\gg \sqrt{t_{min}}$.
From the left panel of Fig.~\ref{dsdtANDdsdtheta} it is seen that the cross-section of ALP production is highly 
suppressed in the region of this value. This means that one can set $t_{max}=\infty$ in the
integration of Eq.~(\ref{dsdt1}) over $t$. Thus the total cross-section of the Primakoff process is
\begin{equation}
\sigma_{tot} \simeq  \frac{1}{2^3} \, g_{a \gamma \gamma}^2 \alpha 
\int\limits_{t_{min}}^{\infty}  \frac{dt}{t^2} (t-t_{min}) 
F^2(t).
\label{IntegralCS1}
\end{equation}
For the  typical wide range of ALP masses, $20\, \mbox{MeV} \lesssim m_a\lesssim 100$ MeV
and typical energies of photons,  $50\, \mbox{GeV} \lesssim E_\gamma \lesssim 100$ GeV, 
the parameters of lead form-factors ($Z=82$ and $A=207$) satisfy  $d \gg t_{min}$ and $d 
\gg 1/a^2$. Given that approach,  one has the following 
expression for the total cross-section in the leading logarithmic order
\begin{equation}
\sigma_{tot}= \frac{16 \pi \alpha}{m_a^3} \cdot \Gamma_{a\rightarrow \gamma \gamma} \cdot \frac{Z^2}{2}
\left(\ln\left[\frac{d}{1/a^2+t_{min}}\right] -2 \right).
\label{TCSofALPprod1}
\end{equation}
The total cross-section depends rather weakly on $m_a$ and $E_\gamma$, see Fig.~\ref{stotVsEg}. Additionaly, in 
Appendix~\ref{AppendixSect} the exact expression for the total 
cross-section Eq.~(\ref{IntegralCS1}) is presented.
One can see from Eq.~(\ref{TotalAnalytical}) that Eq.~(\ref{TCSofALPprod1}) 
is a reasonable approximation of Eq.~(\ref{IntegralCS1}) 
with accuracy better than $1\%$ for the lead form-factor and 
ultra-relativistic ALP with sub-GeV masses.

\begin{figure*}[tbh]
\begin{center}
\includegraphics[width=0.4\textwidth]{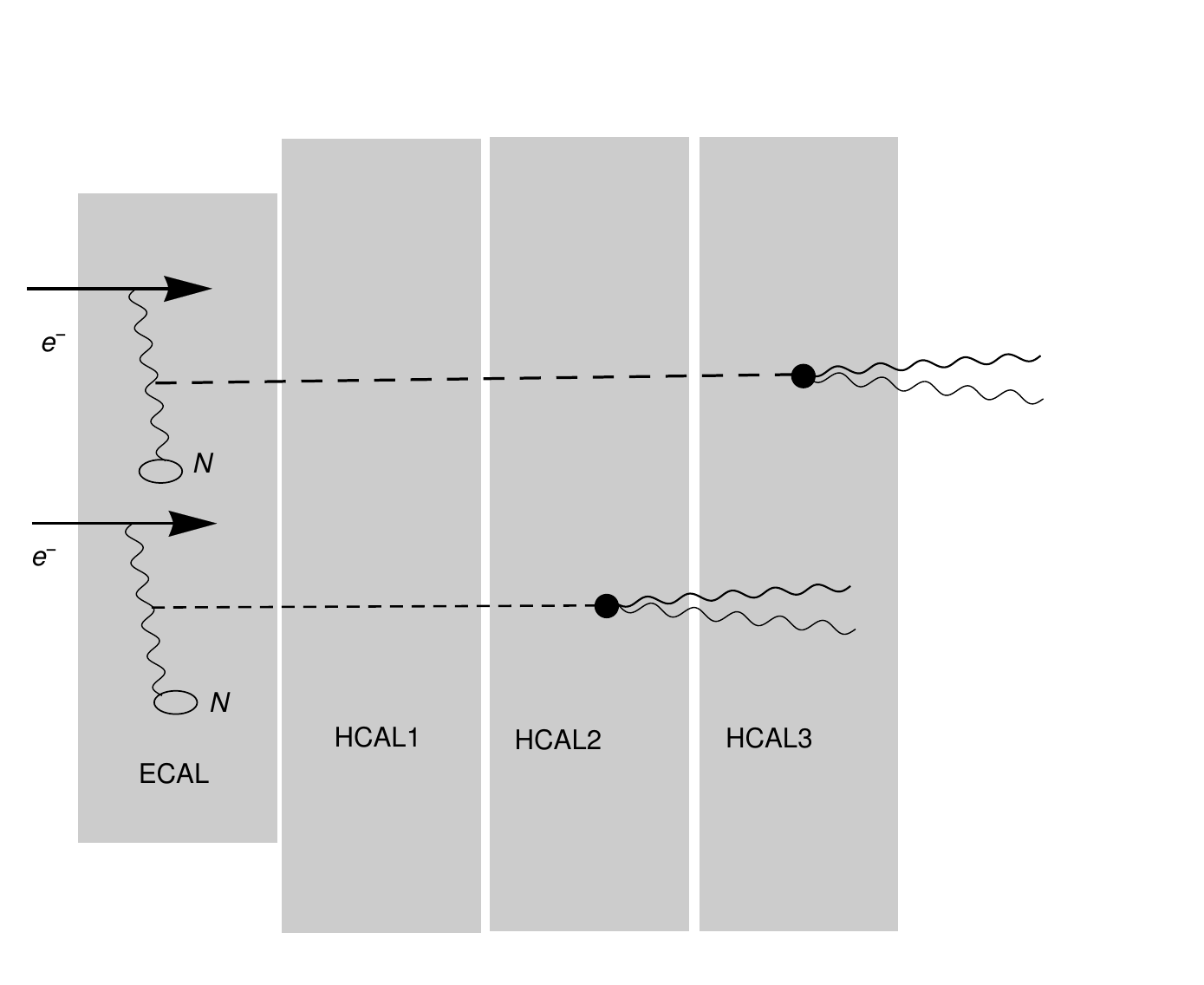}
\includegraphics[width=0.4\textwidth]{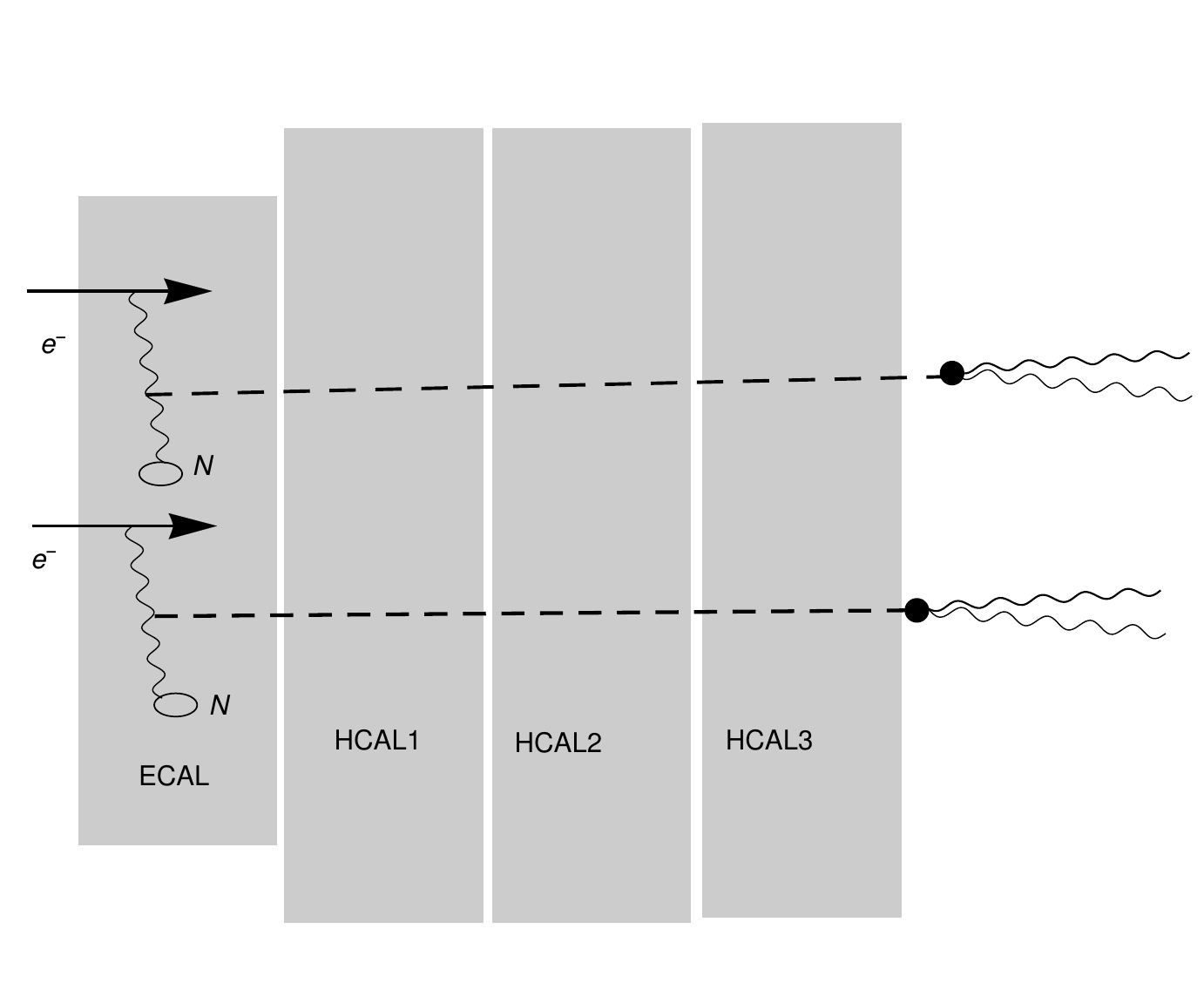}
\caption {The NA64 design for the search for ALP decays,
$a\to \gamma\gamma$. Left panel corresponds to the {\bf Visible Signature}
in NA64, where ALPs decay in the central cells of HCAL2 and HCAL3 
with HCAL1 being a veto. Right panel is the {\bf Invisible Signature} in
NA64, in which we search for decays $a\to \gamma\gamma$ outside all NA64 subdetectors  
\label{figInvDesgn}}
\end{center}
\end{figure*} 

%\subsection{Angular distribution}
From Eqs.~(\ref{DsDEa1}), (\ref{M2viaEa}), (\ref{EaApprox}) and~(\ref{t1def}) we obtain
\begin{equation}
d \sigma \approx \frac{1 }{m_a^3} 16 \pi \alpha F^2(t) \Gamma_{a\rightarrow \gamma \gamma} \frac{\theta_a^3  d\theta_a}{(\theta_a^2+\delta_a^2)^2},  
\end{equation}
where $\delta_a \approx  m_a^2/( 2 E_\gamma^2)$ is a parameter that characterizes a typical angle between the beam line and the ALP momentum.
This result coincides with \cite{Dobrich:2015jyk,Bjorken:1988as,Tsai:1986tx}.
We note that $d \sigma/d \theta_a$ has a peak at
\begin{equation}
\theta^*_a \approx \frac{1}{a E_\gamma} \sqrt{3(1+a^2 t_{min})},
\label{typicalAngle}
\end{equation}
This is a typical angle of ALP, see the right panel of Fig.~\ref{dsdtANDdsdtheta}. For $a^2 t_{min} \gg 1$ it is 
proportional to $\delta_a$, and for  $a^2 t_{min} \ll 1$ it is scaled as 
$\propto 1/(aE_\gamma)$.

\begin{figure}[tbh!]
\begin{center}
\includegraphics[width=0.45\textwidth]{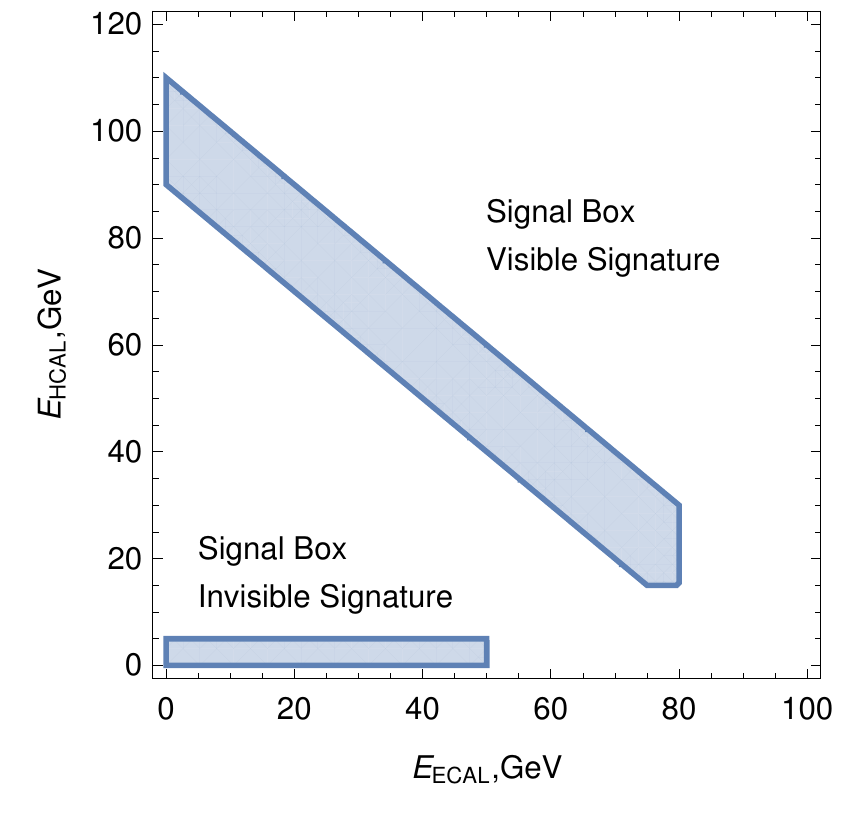}
\caption {Signal Boxes for ALP searches at NA64. 
For the {\bf Visible Signature} it is constructed according to the energy conservation $E_0\simeq E_{ECAL}+E_{HCAL}$ law.
In the {\bf Invisible Signature} a small energy in HCAL and a significant missing energy are required
$E_{miss} = E_0- (E_{ECAL}+E_{HCAL})$.
Here $E_{HCAL}=E_{HCAL1}+E_{HCAL2}+E_{HCAL3}$ is the total energy deposition
in all HCAL modules. For the illustrative purpose we
increase the signal box of {\bf Invisible Signature}
by factor of $5$ along the $E_{HCAL}$ axis.
\label{SignalBoxFigure}}
\end{center}
\end{figure}

\begin{figure}[tbh!]
\begin{center}
\includegraphics[width=0.5\textwidth]{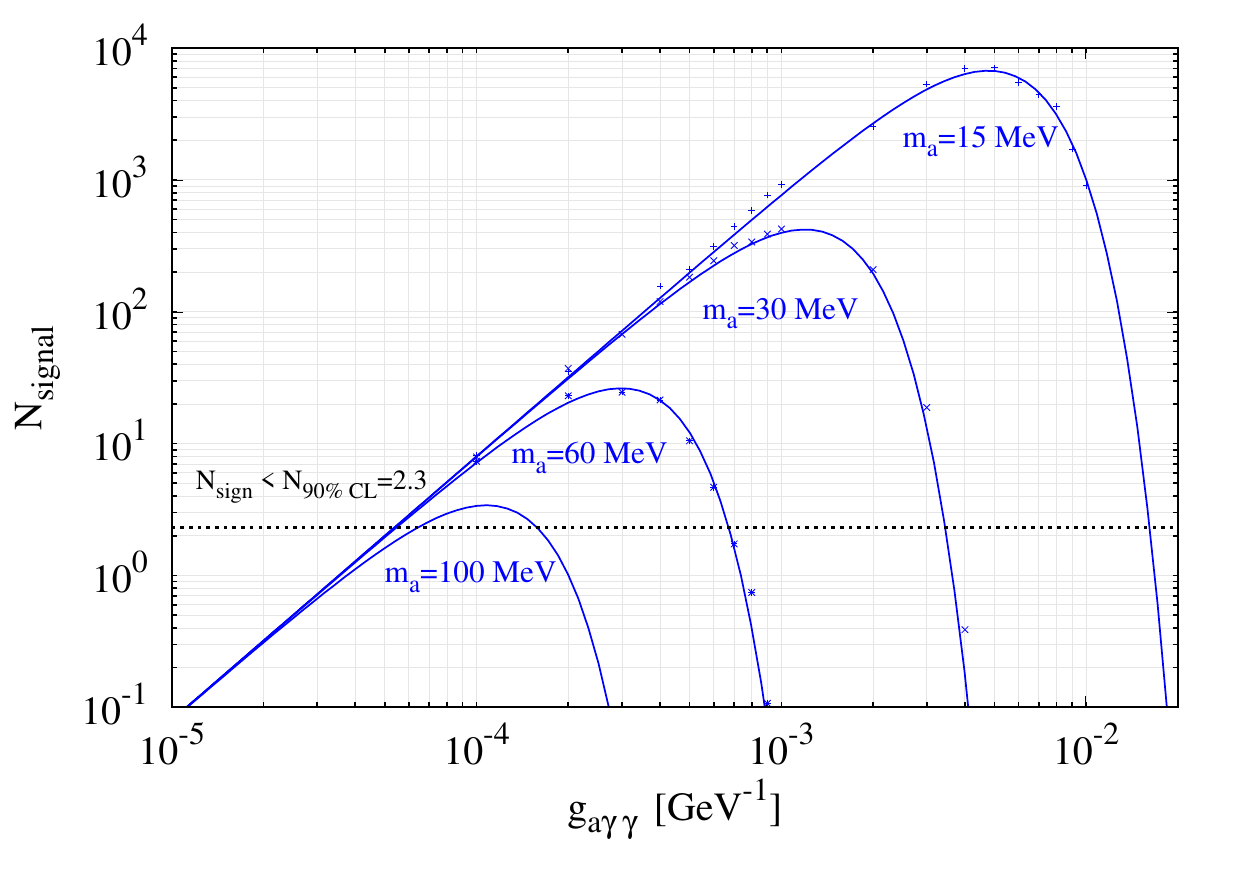}
\caption {Number of signal events as a function of $g_{a\gamma 
\gamma}$ for $N_{EOT}=5\times 10^{12}$.}
\label{InvNsignVsgagg}
\end{center}
\end{figure} 

\begin{figure}[t]
\begin{center}
\includegraphics[width=0.5\textwidth]{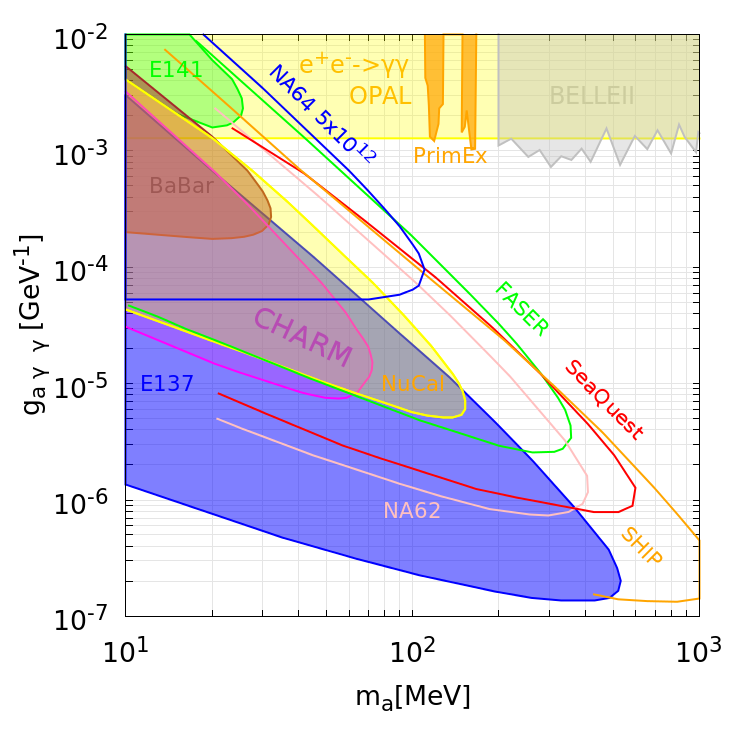}
\caption {The expected 90 \% C.L. sensitivity region of NA64 to the
 ALP production in the Primakoff process, $\gamma N \rightarrow a N$ followed by the
 decay into photon pairs $a \rightarrow \gamma \gamma$ (solid blue line).
 The limits for E137~\cite{Bjorken:1988as}, CHARM~\cite{charm}, NuCal~\cite{nucal}, BaBar~\cite{Aubert:2008as}, 
 E141~\cite{e141b}, LEP~\cite{lep} ($e^+e^-\to \gamma\gamma$) and PrimEx~\cite{primex} are taken from Refs.~\cite{Dolan:2017osp,Dobrich:2019dxc,Knapen:2016moh}.
 Recent bounds on ALPs from {\tt BELLEII}~\cite{BelleII:2020fag} are
  shown by grey region. 
  We also show
 the expected limits for 
 FASER~\cite{Feng:2018noy}, NA62~\cite{Dobrich:2019dxc}, SeaQuest~\cite{Berlin:2018pwi} and SHIP~\cite{Dobrich:2015jyk}.
\label{VisALPfig} }
\end{center}
\end{figure}

\section{Calculation of the ALP yield in NA64
\label{MCSect}}

In this section we discuss the implementation of the code for the MC simulation of the ALP production that uses the formulas derived above
in the full simulation program based on {\tt GEANT4}~\cite{Agostinelli:2002hh} for the NA64 experiment~\cite{Gninenko:2017yus}.

%The photons producing ALP originate from the bremsstrahlung radiation of electrons and positrons within the
%electromagnetic shower at the ECAL detector. 
%That shower is initiated by the incident electrons of $100$~GeV
%and predominantly absorbed within the ECAL itself.  
%The Primakoff process of ALP production $\gamma N \rightarrow a N$
%is probed for every photon with energy $E_\gamma$ above some
%threshold corresponding to minimal ALP energy that the NA64 facility is
%capable to detect. The SM processes enrolled in standard physics list
%is simulated as well.

The photons that can produce ALP originate from the bremsstrahlung radiation of electrons and positrons of the
electromagnetic shower from the primary 100 GeV electron beam absorbed in the target - calorimeter ECAL. 
The Primakoff process of ALP production $\gamma N \rightarrow a N$ in this program can occur along with other, SM processes,
for all photons of the electromagnetic shower if the photon energy $E_{\gamma}$ is above some threshold that corresponds
to the minimal detectable ALP energy.

Now we describe the calculation of the ALP signal in NA64 at each step of the photon propagation in the target.
The number of ALP produced at the $i$-th photon's step in the electromagnetic shower is
\begin{equation}
N_{a}^{(i)} = \frac{\rho N_A}{A} \sigma_{tot}(E_\gamma^i) \times  l_i
\end{equation}
where $\rho$ is a density of the
lead ECAL active target, $A$ is the atomic mass of the target, $N_A$ is
Avogadro's number, $\sigma_{tot}(E_\gamma^i)$ is the total
cross-section of ALP elastic interaction with a nucleus (see, e.~g.
Eq.~(\ref{TCSofALPprod1}) for details), $l_i$ is the step length of the 
photon in target. 

%In the {\tt GEANT4} simulation, 
%for each step of track propagation the following accept/reject
%     scheme is applied to simulate a set of signal samples:
In the simulation of signal samples, at each step of tracing of a photon
with the energy above threshold the following actions (accept/reject scheme) are made:

\begin{itemize}
\item
we randomly sample the variable $u$ distributed
uniformly in the range $[0,1]$, if $u$ is smaller than the $a$ emission probability
$$
P_{emission} = \frac{\rho N_A}{A} \times \sigma_{tot} (E_\gamma^i) \times  l_i
$$
then the emission of $a$ is accepted,
\item
for each emitted $a$ we then generate the value of $E_a/E_\gamma$ and the angle of $a$ w.r.t. the initial photon
according to the differential cross section (Fig.~\ref{dsdtANDdsdtheta}), then we calculate the four-momentum of ALP.
The value of $E_a/E_\gamma$ is very close to unity, $E_a/E_\gamma \simeq 1$,
\item  after production the ALP decay is simulated according to Eq.(\ref{Width}).
\end{itemize}

 In order to simulate samples with sufficient total statistics we used the CERN batch system.
This production process was automatized \cite{RenatGit}.

\section{The ALP search strategy  
\label{Strategy}}

We assumed the configuration of NA64 \cite{Banerjee:2019invis} initially designed for the search for invisible decays of dark photons $A'$,
which is suitable as well for the ALP search, see Fig.~\ref{figInvDesgnFull}. In this configuration the target - calorimeter ECAL
is followed by three modules of the hadron calorimeter HCAL. The 100 GeV beam of electrons is cleaned from other particles by the two magnets
MBPL and the synchrotron radiation detector SRD to the level of $10^{-6}$. The momentum of the incident
electrons is measured with accuracy $\simeq 1\%$.
%We assume that ALPs decay predominantly to photons $\mathcal{B}(a\rightarrow \gamma \gamma)\approx 1$.

 Two distinct signatures of ALP in the NA64 experiment are possible.
In the first  signature ({\bf Visible Signature} in Fig.~\ref{figInvDesgn}) the ALP decays in the second and third modules of HCAL, the first module (HCAL1) serving as a veto.
So we required the energy deposition compatible with noise in HCAL1 (below 1 GeV) and at least 15 GeV in HCAL2 and HCAL3.
%In addition, the energy deposition distributions in HCAL2 and HCAL3
%should be compatible with two nearly collinear photons from the
%ALP decay $a \rightarrow \gamma \gamma$. The produced photon pair is tending
%to be emitted closely to the incident particle direction because of
%the ALP emission angle itself being small. 
In addition, the energy deposition distributions in HCAL2 and HCAL3 
should be compatible with two nearly collinear photons from the ALP decay $a\to \gamma \gamma$,
very close to the electron beam axis because of very smal angles of the ALP emission.
This means that almost all energy (more than 95\%) should
be deposited in the central cell of the HCAL modules. This is important for the background suppression
because hadronic showers are usually much wider and deposit significant energy in peripheric cells of HCAL.
We also require that the energy is conserved, taking into account the energy resolution of calorimeters:
\begin{equation}
|E_{ECAL} +E_{HCAL} - E_0| \lesssim 10\,~\mbox{GeV},
\label{VisSetupECL}
\end{equation}
where $E_{ECAL}$ and $E_{HCAL}$ are the energy depositions in electromagnetic and hadronic calorimeters respectively,
$E_0\simeq 100$~GeV is the energy of primary electrons.
For this signature we don't apply any additional cut on $E_{ECAL}$, accepting all events that passed the normal hardware trigger
cut used in NA64 in all physical runs,
\begin{equation}
E_{ECAL} \lesssim 80\,~\mbox{GeV}.
\end{equation}
The corresponding signal box is shown in Fig.~\ref{SignalBoxFigure}.

In the second signature ({\bf Invisible Signature}) the ALP decays beyond all subdetectors of NA64. This is the missing energy signature of ALP,
the same that is described in 
Ref.~\cite{Banerjee:2016tad,Banerjee:2019invis}.
%$e^-N\to e^- N a \to e^- N +E_{miss}$. 
The selection criteria can be found in the corresponding references. 
The most important cuts are the upper ECAL
energy cut and the requirement of no energy deposition in all three HCAL modules HCAL1 - HCAL3,
\begin{equation}
E_{ECAL} \lesssim 50\,~\mbox{GeV}, \qquad E_{HCAL} \lesssim 1\,~\mbox{GeV}.
\end{equation}
In the missing energy signature the cut on the energy in the ECAL is rather strict
This means that only shower photons with the energy above~$50$ GeV can produce
detectable ALP. This signature is shown in the right panel of Fig.~\ref{figInvDesgn}.
The corresponding signal box is shown in Fig.~\ref{SignalBoxFigure}. 

Note that the background conditions in the {\bf Visible Signature} are much better, for this reason it was possible to relax the
cut on $E_{ECAL}$ to 80 GeV as compared to the invisible one. Correspondingly, in this signature photons
with the energy as low as $\simeq 20$ GeV can produce detectable ALP. The flux of such photons is significantly higher than
thouse above 50 GeV, detectable in the invisible signature. For this reason the contribution of the {\bf Visible Signature}
to the total sensitivity of NA64 to ALPs is significant.
In the signal samples we simulated the ALP with the energy $E_{ALP} > 18$ GeV decaying beyond the HCAL1 module, which includes also
$a\to \gamma\gamma$ decays far from the NA64 detectors. The cuts corresponding to the two signatures were applied during the
processing of these samples by the reconstruction program.

%The background analysis for ALP signal was carried out in 
%Ref.~\cite{Gninenko:2018ter} for the physical runs of NA64
%with $2.84\times 10^{11}$  electrons on target. 
%The processes, which can mimic the ALP signal in detectors 
%are 
%\begin{itemize}
%\item the production of a leading neutrons in HCAL modules,
%\item the reaction of $K_0$ meson production in ECAL,
%$e N \to n(K_0)+m\pi_0 +X$,
%\item electron beam contamination due to $\pi^-$ and $K^-$,
%\item dimuon production in ECAL, $e^- N \to e^- N \mu^+\mu^-$.
%\end{itemize}
% As shown in \cite{Gninenko:2018ter}, the background in both signatures is smaller than $0.14$ events for the
%corresponding statistics of $2.84\times 10^{11}$ EOT. 
%The increase of statistics to $5\times 10^{12}$ EOT will require 
%the upgrade of the NA64 detector in 2020 - 2021, which should significantly suppress the background.

 The background in the {\bf Visible Signature} is caused mainly by the punch-through leading $K_0$ and neutrons produced
in electronuclear interactions in ECAL \cite{Banerjee:2020fue}. The background in both signatures is shown to be smaller
than $0.2$ events \cite{Banerjee:2020fue,Banerjee:2019invis}, so that it can be neglected in the sensitivity estimates
since the difference from the background free case is small. After the upgrade of the NA64 detector in 2020 - 2021
it will be further suppressed.

 In the real experiment the simultaneous statistical analysis of the both signatures is to be performed. However, for the
sensitivity estimation in the conditions of small background we can simply sum up the numbers of expected signal events 
for {\bf Visible} and {\bf Invisible} signatures. 

\section{The expected sensitivity of NA64
\label{ExpectedlimitsNA64}}
 Now we estimate the sensitivity of NA64 to ALPs.
The number of detectable ALPs can be written as
\begin{equation}
N_{a} = \frac{N_{EOT}}{N_{MC}} \sum_i  N_a^{(i)}
\exp\left(-L_{D}^{(i)}/l^{(i)}_a\right) 
\mathcal{B}(a \rightarrow \gamma \gamma),
\label{visYield2}
\end{equation}
where $N_{EOT}$ is the number of electrons on target in the experiment, $N_{MC}$ is the number of simulated events,
%$N_a^{(i)}$ is the ALP yield from the $(i)$-th step of production photon in the target,
$L_D^{(i)}$ is the distance from the production point to the minimal 
allowed decay point coordinate $Z_{min}$,
$l^{(i)}_a$ is the ALP decay length taking into account its Lorentz factor, see, e.~g.~Eq.~(\ref{DecLength1}). 
$Z_{min}$ can be the end of HCAL1 or the end of HCAL3 depending
on the signature under study. The typical lengths here are the lengths of the calorimeters $L_{ECAL}$=45cm
and $L_{HCAL~module}$=1.3m. The typical energy of ALP in the Primakoff process is
$E_a \approx E_\gamma$, therefore the ALP spectra are associated with the spectra of shower photons in the dump.

 In Fig.~\ref{InvNsignVsgagg} we show the number of 
 $a\to \gamma\gamma$ decays as a
 function of ALP coupling with photons. Assuming background free case and zero 
 signal events observed at NA64 we require $90\% CL$ upper limit on the number
 of ALP decays to be $N_{90\%}=2.3$ according to the Poisson statistics.
 For each ALP mass $m_a$ the range of couplings constrained
$g_{a\gamma\gamma}^{low}(m_a)<g_{a\gamma\gamma}<g_{a\gamma\gamma}^{up}(m_a)$
is defined by inequality $N_a>N_{90\%}$, see Fig.~\ref{InvNsignVsgagg}, 
The values above $g_{a\gamma\gamma}^{up}$ correspond to short-lived ALP decaying prematurely, before reaching the veto.
The values below $g_{a\gamma\gamma}^{low}$ correspond to too small signal yield.
The resulting plot in Fig.~\ref{InvNsignVsgagg} includes both visible and invisible signatures.

In Fig.~\ref{VisALPfig} we show the $90\%$ C.L. sensitivity region of the NA64 experiment
in the ALP parameters space for the background free case and the total number of $100$~GeV electrons on target
$N_{EOT}=5\times 10^{12}$ in the mass range $10\, \mbox{MeV}\lesssim m_a \lesssim 100$~MeV.
The additional inefficiency of the detector due to instrumental effects not included in the
MC-simulation is assumed to be negligible. It was below $20\%$ in the published relult \cite{Banerjee:2020fue} and will be significantly
decreased after the detector upgrade in 2020 - 2021.

These results demonstrate 
that the NA64 experiment is capable to probe the
ALP coupling with photons in the range $5\times 10^{-5}\, \mbox{GeV}^{-1} \lesssim g_{a\gamma\gamma} \lesssim 10^{-3}\, \mbox{GeV}^{-1}$.

\section{Conclusion
\label{Concl}}
  
 In the present manuscript we have studied the prospects of the fixed target experiment NA64 that uses the electron beam at the CERN SPS
to search for axion-like particles.
In particular, we have studied the properties of the ALP production in the Primakoff reaction $\gamma N\to N a$ and its decay.
We have implemented the ALP production cross-sections and the process of its decay in the NA64 simulation program
based on the {\tt GEANT4} toolkit.
We have calculated the expected sensitivity to ALP of the NA64 experiment and have shown that it
potentially allows to examine the unexplored region in the parameter space $5\times 10^{-5}\, \mbox{GeV}^{-1} \lesssim g_{a\gamma\gamma} \lesssim 10^{-3}\,
\mbox{GeV}^{-1}$ and $10\, \mbox{MeV} \lesssim m_a\lesssim 100$ MeV if the statistics corresponding to $N_{EOT}=5\times 10^{12}$
electrons on target is accumulated.

\section{Acknowledgements}
We would like to thank our colleagues from the NA64 Collaboration, in particular, 
S.~N.~Gninenko, N.~V.~Krasnikov, P.~Crivelli, D.~S.~Gorbunov and 
V.~E.~Lyubovitskij for useful discussions. 
This work was supported by the Ministry of Science and Higher Education (MSHE) in the frame of the Agreement on 23 July 2020
No 075-15-2020-718 ID No RFMEFI61320X0098.

\appendix
\section{Exact formula for the total cross-section
\label{AppendixSect}}
The full analytical expression for the 
integral (\ref{IntegralCS1}) is
$$
\sigma_{tot} = \frac{16 \pi^2}{m_a^3} 
\Gamma_{a\to \gamma\gamma} \frac{Z^2}{2} \frac{d^2}{(d-1/a^2)^3} \times
$$
\begin{equation}
\left[ (d+2t_{min}+1/a^2) \log \left(\frac{d+t_{min}}{1/a^2+t_{min}}\right) -2 d+2/a^2 \right].
\label{TotalAnalytical}
\end{equation}
For the parameter space of interest, $d\gg 1/a^2$ and $d\gg t_{min}$ the difference between 
the analytical~(\ref{TotalAnalytical}) and approximate expressions (\ref{TCSofALPprod1}) is below $1\%$.


\begin{thebibliography}{99}


%\cite{Svrcek:2006yi}
\bibitem{Svrcek:2006yi}
  P.~Svrcek and E.~Witten,
  %``Axions In String Theory,''
  JHEP {\bf 0606} (2006) 051
 % doi:10.1088/1126-6708/2006/06/051
  [hep-th/0605206].
  %%CITATION = doi:10.1088/1126-6708/2006/06/051;%%
  %772 citations counted in INSPIRE as of 28 Mar 2020


%\cite{Visinelli:2018utg}
\bibitem{Visinelli:2018utg}
L.~Visinelli and S.~Vagnozzi,
%``Cosmological window onto the string axiverse and the supersymmetry breaking scale,''
Phys. Rev. D \textbf{99} (2019) no.6, 063517
%doi:10.1103/PhysRevD.99.063517
[arXiv:1809.06382 [hep-ph]].
%43 citations counted in INSPIRE as of 17 Aug 2020

%\cite{Arvanitaki:2009fg}
\bibitem{Arvanitaki:2009fg}
  A.~Arvanitaki, S.~Dimopoulos, S.~Dubovsky, N.~Kaloper and J.~March-Russell,
  %``String Axiverse,''
  Phys.\ Rev.\ D {\bf 81} (2010) 123530
 % doi:10.1103/PhysRevD.81.123530
  [arXiv:0905.4720 [hep-th]].
  %%CITATION = doi:10.1103/PhysRevD.81.123530;%%
  %763 citations counted in INSPIRE as of 28 Mar 2020


%\cite{Nelson:1993nf}
\bibitem{Nelson:1993nf}
  A.~E.~Nelson and N.~Seiberg,
  %``R symmetry breaking versus supersymmetry breaking,''
  Nucl.\ Phys.\ B {\bf 416} (1994) 46
 % doi:10.1016/0550-3213(94)90577-0
  [hep-ph/9309299].
  %%CITATION = doi:10.1016/0550-3213(94)90577-0;%%
  %341 citations counted in INSPIRE as of 28 Mar 2020

%\cite{Bagger:1994hh}
\bibitem{Bagger:1994hh}
  J.~Bagger, E.~Poppitz and L.~Randall,
  %``The R axion from dynamical supersymmetry breaking,''
  Nucl.\ Phys.\ B {\bf 426} (1994) 3
%  doi:10.1016/0550-3213(94)90123-6
  [hep-ph/9405345].
  %%CITATION = doi:10.1016/0550-3213(94)90123-6;%%
  %173 citations counted in INSPIRE as of 28 Mar 2020

%\cite{Peccei:1977hh}
\bibitem{Peccei:1977hh}
  R.~D.~Peccei and H.~R.~Quinn,
  %``CP Conservation in the Presence of Instantons,''
  Phys.\ Rev.\ Lett.\  {\bf 38} (1977) 1440.
 % doi:10.1103/PhysRevLett.38.1440
  %%CITATION = doi:10.1103/PhysRevLett.38.1440;%%
  %5186 citations counted in INSPIRE as of 28 Mar 2020



%\cite{Weinberg:1977ma}
\bibitem{Weinberg:1977ma}
  S.~Weinberg,
  %``A New Light Boson?,''
  Phys.\ Rev.\ Lett.\  {\bf 40} (1978) 223.
 % doi:10.1103/PhysRevLett.40.223
  %%CITATION = doi:10.1103/PhysRevLett.40.223;%%
  %3590 citations counted in INSPIRE as of 28 Mar 2020
  
  %\cite{Wilczek:1977pj}
\bibitem{Wilczek:1977pj}
  F.~Wilczek,
  %``Problem of Strong  $P$  and  $T$  Invariance in the Presence of Instantons,''
  Phys.\ Rev.\ Lett.\  {\bf 40} (1978) 279.
%  doi:10.1103/PhysRevLett.40.279
  %%CITATION = doi:10.1103/PhysRevLett.40.279;%%
  %3446 citations counted in INSPIRE as of 28 Mar 2020
  

  
%\cite{Boehm:2003hm}
\bibitem{Boehm:2003hm}
  C.~Boehm and P.~Fayet,
  %``Scalar dark matter candidates,''
  Nucl.\ Phys.\ B {\bf 683} (2004) 219
  doi:10.1016/j.nuclphysb.2004.01.015
  [hep-ph/0305261].
  %%CITATION = doi:10.1016/j.nuclphysb.2004.01.015;%%
  %481 citations counted in INSPIRE as of 04 Apr 2020  
  
  
%\cite{Dolan:2014ska}
\bibitem{Dolan:2014ska}
  M.~J.~Dolan, F.~Kahlhoefer, C.~McCabe and K.~Schmidt-Hoberg,
  %``A taste of dark matter: Flavour constraints on pseudoscalar mediators,''
  JHEP {\bf 1503} (2015) 171
   Erratum: [JHEP {\bf 1507} (2015) 103]
 % doi:10.1007/JHEP07(2015)103, 10.1007/JHEP03(2015)171
  [arXiv:1412.5174 [hep-ph]].
  %%CITATION = doi:10.1007/JHEP07(2015)103, 10.1007/JHEP03(2015)171;%%
  %98 citations counted in INSPIRE as of 28 May 2018  
  
%\cite{Hochberg:2018rjs}
\bibitem{Hochberg:2018rjs}
Y.~Hochberg, E.~Kuflik, R.~Mcgehee, H.~Murayama and K.~Schutz,
%``Strongly interacting massive particles through the axion portal,''
Phys. Rev. D \textbf{98} (2018) no.11, 115031
doi:10.1103/PhysRevD.98.115031
[arXiv:1806.10139 [hep-ph]].
%36 citations counted in INSPIRE as of 17 Aug 2020  
  

%\cite{Gninenko:2016kpg}
\bibitem{Gninenko:2016kpg}
  S.~N.~Gninenko, N.~V.~Krasnikov, M.~M.~Kirsanov and D.~V.~Kirpichnikov,
  %``Missing energy signature from invisible decays of dark photons at the CERN SPS,''
  Phys.\ Rev.\ D {\bf 94} (2016) no.9,  095025
 % doi:10.1103/PhysRevD.94.095025
  [arXiv:1604.08432 [hep-ph]].
  %%CITATION = doi:10.1103/PhysRevD.94.095025;%%
  %19 citations counted in INSPIRE as of 11 Feb 2019

%\cite{Banerjee:2016tad}
\bibitem{Banerjee:2016tad}
  D.~Banerjee {\it et al.} [NA64 Collaboration],
  %``Search for invisible decays of sub-GeV dark photons in missing-energy events at the CERN SPS,''
  Phys.\ Rev.\ Lett.\  {\bf 118} (2017) no.1,  011802
%  doi:10.1103/PhysRevLett.118.011802
  [arXiv:1610.02988 [hep-ex]].
  %%CITATION = doi:10.1103/PhysRevLett.118.011802;%%
  %67 citations counted in INSPIRE as of 11 Feb 2019
  
%\cite{Gninenko:2017yus}
\bibitem{Gninenko:2017yus}
  S.~N.~Gninenko, D.~V.~Kirpichnikov, M.~M.~Kirsanov and N.~V.~Krasnikov,
  %``The exact tree-level calculation of the dark photon production in high-energy electron scattering at the CERN SPS,''
  Phys.\ Lett.\ B {\bf 782} (2018) 406
%  doi:10.1016/j.physletb.2018.05.010
  [arXiv:1712.05706 [hep-ph]].
  %%CITATION = doi:10.1016/j.physletb.2018.05.010;%%
  %8 citations counted in INSPIRE as of 11 Feb 2019
  
%\cite{Gninenko:2018ter}
\bibitem{Gninenko:2018ter}
  S.~N.~Gninenko, D.~V.~Kirpichnikov and N.~V.~Krasnikov,
  %``Probing millicharged particles with NA64 experiment at CERN,''
  Phys.\ Rev.\ D {\bf 100} (2019) no.3,  035003
 % doi:10.1103/PhysRevD.100.035003
  [arXiv:1810.06856 [hep-ph]].
  %%CITATION = doi:10.1103/PhysRevD.100.035003;%%
  %9 citations counted in INSPIRE as of 07 Apr 2020

\bibitem{Banerjee:2019invis}
  D.~Banerjee {\it et al.} [NA64 Collaboration],
  %``Dark matter search in missing energy events with NA64,''
  Phys.\ Rev.\ Lett.\  {\bf 123} (2019) no.12, 121801
%  doi:10.1103/PhysRevLett.123.121801
  [arXiv:1906.00176 [hep-ex]].

  %\cite{Krasznahorkay:2015iga}
%\bibitem{Krasznahorkay:2015iga}
%  A.~J.~Krasznahorkay {\it et al.},
  %``Observation of Anomalous Internal Pair Creation in Be8 : A Possible Indication of a Light, Neutral Boson,''
%  Phys.\ Rev.\ Lett.\  {\bf 116} (2016) no.4,  042501
 % doi:10.1103/PhysRevLett.116.042501
%  [arXiv:1504.01527 [nucl-ex]].
  %%CITATION = doi:10.1103/PhysRevLett.116.042501;%%
  %104 citations counted in INSPIRE as of 12 Feb 2019
  
  
%\cite{Banerjee:2018vgk}
%\bibitem{Banerjee:2018vgk}
%  D.~Banerjee {\it et al.} [NA64 Collaboration],
  %``Search for a Hypothetical 16.7 MeV Gauge Boson and Dark Photons in the NA64 Experiment at CERN,''
%  Phys.\ Rev.\ Lett.\  {\bf 120} (2018) no.23,  231802
 % doi:10.1103/PhysRevLett.120.231802
%  [arXiv:1803.07748 [hep-ex]].
  %%CITATION = doi:10.1103/PhysRevLett.120.231802;%%
  %13 citations counted in INSPIRE as of 11 Feb 2019

%\cite{Banerjee:2019hmi}
%\bibitem{Banerjee:2019hmi}
%  D.~Banerjee {\it et al.} [NA64 Collaboration],
  %``Improved limits on a hypothetical X(16.7) boson and a dark photon decaying into $e^+e^-$ pairs,''
%  arXiv:1912.11389 [hep-ex].
  %%CITATION = ARXIV:1912.11389;%%
  %6 citations counted in INSPIRE as of 07 Apr 2020

%\cite{Ellwanger:2016wfe}
%\bibitem{Ellwanger:2016wfe}
%U.~Ellwanger and S.~Moretti,
%``Possible Explanation of the Electron Positron Anomaly at 17 MeV in $^8Be$ Transitions Through a Light Pseudoscalar,''
%JHEP \textbf{11} (2016), 039
%doi:10.1007/JHEP11(2016)039
%[arXiv:1609.01669 [hep-ph]].
%48 citations counted in INSPIRE as of 04 Aug 2020

%\cite{Batell:2016ove}
%\bibitem{Batell:2016ove}
%  B.~Batell, N.~Lange, D.~McKeen, M.~Pospelov, and A.~Ritz,
  %``Muon anomalous magnetic moment through the leptonic Higgs portal,''
%  Phys.\ Rev.\ D {\bf 95},  075003  (2017)
%  [arXiv:1606.04943 [hep-ph]].

%\cite{Chen:2015vqy}
%\bibitem{Chen:2015vqy}
%  C.~Y.~Chen, H.~Davoudiasl, W.~J.~Marciano, and C.~Zhang,
  %``Implications of a light “dark Higgs” solution to the $g_μ$-2 discrepancy,''
%  Phys.\ Rev.\ D {\bf 93}, 035006 (2016)
%  [arXiv:1511.04715 [hep-ph]].

%\bibitem{KirpPseudoScal}  Kirpichnikov~D.~V. et al. in preparation. 

%\cite{Ariga:2018uku}
\bibitem{Ariga:2018uku}
  A.~Ariga {\it et al.} [FASER Collaboration],
  %``FASER’s physics reach for long-lived particles,''
  Phys.\ Rev.\ D {\bf 99} (2019) no.9,  095011
 % doi:10.1103/PhysRevD.99.095011
  [arXiv:1811.12522 [hep-ph]].
  %%CITATION = doi:10.1103/PhysRevD.99.095011;%%
  %49 citations counted in INSPIRE as of 07 Apr 2020
  
  %\cite{Chou:2016lxi}
\bibitem{Chou:2016lxi}
  J.~P.~Chou, D.~Curtin and H.~J.~Lubatti,
  %``New Detectors to Explore the Lifetime Frontier,''
  Phys.\ Lett.\ B {\bf 767} (2017) 29
 % doi:10.1016/j.physletb.2017.01.043
  [arXiv:1606.06298 [hep-ph]].
  %%CITATION = doi:10.1016/j.physletb.2017.01.043;%%
  %87 citations counted in INSPIRE as of 08 Feb 2019


%\cite{Anelli:2015pba}
\bibitem{Anelli:2015pba}
  M.~Anelli {\it et al.} [SHiP Collaboration],
  %``A facility to Search for Hidden Particles (SHiP) at the CERN SPS,''
  arXiv:1504.04956 [physics.ins-det].
  %%CITATION = ARXIV:1504.04956;%%
  %201 citations counted in INSPIRE as of 08 Feb 2019

%\cite{Gligorov:2017nwh}
\bibitem{Gligorov:2017nwh}
  V.~V.~Gligorov, S.~Knapen, M.~Papucci and D.~J.~Robinson,
  %``Searching for Long-lived Particles: A Compact Detector for Exotics at LHCb,''
  Phys.\ Rev.\ D {\bf 97} (2018) no.1,  015023
 % doi:10.1103/PhysRevD.97.015023
  [arXiv:1708.09395 [hep-ph]].
  %%CITATION = doi:10.1103/PhysRevD.97.015023;%%
  %54 citations counted in INSPIRE as of 08 Feb 2019  


%\cite{Berlin:2018pwi}
\bibitem{Berlin:2018pwi}
  A.~Berlin, S.~Gori, P.~Schuster and N.~Toro,
  %``Dark Sectors at the Fermilab SeaQuest Experiment,''
  Phys.\ Rev.\ D {\bf 98} (2018) no.3,  035011
 % doi:10.1103/PhysRevD.98.035011
  [arXiv:1804.00661 [hep-ph]].
  %%CITATION = doi:10.1103/PhysRevD.98.035011;%%
  %22 citations counted in INSPIRE as of 08 Feb 2019

%\cite{Akesson:2018vlm}
\bibitem{Akesson:2018vlm}
  T.~Akesson {\it et al.} [LDMX Collaboration],
  %``Light Dark Matter eXperiment (LDMX),''
  arXiv:1808.05219 [hep-ex].
  %%CITATION = ARXIV:1808.05219;%%
  %10 citations counted in INSPIRE as of 08 Feb 2019

%\cite{Beacham:2019nyx}
\bibitem{Beacham:2019nyx}
  J.~Beacham {\it et al.},
  %``Physics Beyond Colliders at CERN: Beyond the Standard Model Working Group Report,''
  J.\ Phys.\ G {\bf 47} (2020) no.1,  010501
 % doi:10.1088/1361-6471/ab4cd2
  [arXiv:1901.09966 [hep-ex]].
  %%CITATION = doi:10.1088/1361-6471/ab4cd2;%%
  %77 citations counted in INSPIRE as of 07 Apr 2020

%\cite{Alemany:2019vsk}
\bibitem{Alemany:2019vsk}
  R.~Alemany {\it et al.},
  %``Summary Report of Physics Beyond Colliders at CERN,''
  arXiv:1902.00260 [hep-ex].
  %%CITATION = ARXIV:1902.00260;%%

%\cite{Dobrich:2015jyk}
\bibitem{Dobrich:2015jyk}
  B.~Dobrich, J.~Jaeckel, F.~Kahlhoefer, A.~Ringwald and K.~Schmidt-Hoberg,
  %``ALPtraum: ALP production in proton beam dump experiments,''
  JHEP {\bf 1602} (2016) 018, 
 %  [JHEP {\bf 1602} (2016) 018]
 % doi:10.1007/JHEP02(2016)018
  [arXiv:1512.03069~[hep-ph]].
  %%CITATION = doi:10.1007/JHEP02(2016)018;%%
  %35 citations counted in INSPIRE as of 13 May 2018
  
%\cite{Bjorken:1988as}
\bibitem{Bjorken:1988as}
  J.~D.~Bjorken {\it et al.},
  %``Search for Neutral Metastable Penetrating Particles Produced in the SLAC Beam Dump,''
  Phys.\ Rev.\ D {\bf 38} (1988) 3375.
 % doi:10.1103/PhysRevD.38.3375
  %%CITATION = doi:10.1103/PhysRevD.38.3375;%%
  %164 citations counted in INSPIRE as of 20 Jun 2018
  
%\cite{Tsai:1986tx}
\bibitem{Tsai:1986tx}
  Y.~S.~Tsai,
  %``Axion Bremsstrahlung By An Electron Beam,''
  Phys.\ Rev.\ D {\bf 34} (1986) 1326.
%  doi:10.1103/PhysRevD.34.1326
  %%CITATION = doi:10.1103/PhysRevD.34.1326;%%
  %44 citations counted in INSPIRE as of 13 May 2018
  
  
%\cite{Liu:2017htz}
\bibitem{Liu:2017htz}
  Y.~S.~Liu and G.~A.~Miller,
  %``Validity of the Weizsäcker-Williams approximation and the analysis of beam dump experiments: Production of an axion, a dark photon, or a new axial-vector boson,''
  Phys.\ Rev.\ D {\bf 96} (2017) no.1,  016004
 % doi:10.1103/PhysRevD.96.016004
  [arXiv:1705.01633 [hep-ph]].
  %%CITATION = doi:10.1103/PhysRevD.96.016004;%%
  %12 citations counted in INSPIRE as of 22 Jan 2019

%\cite{Liu:2016mqv}
\bibitem{Liu:2016mqv}
  Y.~S.~Liu, D.~McKeen and G.~A.~Miller,
  %``Validity of the Weizsäcker-Williams approximation and the analysis of beam dump experiments: Production of a new scalar boson,''
  Phys.\ Rev.\ D {\bf 95} (2017) no.3,  036010
%  doi:10.1103/PhysRevD.95.036010
  [arXiv:1609.06781 [hep-ph]].
  %%CITATION = doi:10.1103/PhysRevD.95.036010;%%
  %14 citations counted in INSPIRE as of 22 Jan 2019
  
%\cite{Beranek:2013nqa}
\bibitem{Beranek:2013nqa}
  T.~Beranek and M.~Vanderhaeghen,
  %``Study of the discovery potential for hidden photon emission at future electron scattering fixed target experiments,''
  Phys.\ Rev.\ D {\bf 89} (2014) no.5,  055006
 % doi:10.1103/PhysRevD.89.055006
  [arXiv:1311.5104 [hep-ph]].
  %%CITATION = doi:10.1103/PhysRevD.89.055006;%%
  %11 citations counted in INSPIRE as of 22 Jan 2019    
  
%\cite{Beranek:2013yqa}
\bibitem{Beranek:2013yqa}
  T.~Beranek, H.~Merkel and M.~Vanderhaeghen,
  %``Theoretical framework to analyze searches for hidden light gauge bosons in electron scattering fixed target experiments,''
  Phys.\ Rev.\ D {\bf 88} (2013) 015032
%  doi:10.1103/PhysRevD.88.015032
  [arXiv:1303.2540 [hep-ph]].
  %%CITATION = doi:10.1103/PhysRevD.88.015032;%%
  %45 citations counted in INSPIRE as of 22 Jan 2019  
  
%\cite{Bjorken:2009mm}
\bibitem{Bjorken:2009mm}
  J.~D.~Bjorken, R.~Essig, P.~Schuster and N.~Toro,
  %``New Fixed-Target Experiments to Search for Dark Gauge Forces,''
  Phys.\ Rev.\ D {\bf 80} (2009) 075018
 % doi:10.1103/PhysRevD.80.075018
  [arXiv:0906.0580 [hep-ph]].
  %%CITATION = doi:10.1103/PhysRevD.80.075018;%%
  %545 citations counted in INSPIRE as of 11 Aug 2020  
  
 %\cite{Shtabovenko:2016sxi}
\bibitem{Shtabovenko:2016sxi}
  V.~Shtabovenko, R.~Mertig and F.~Orellana,
  %``New Developments in FeynCalc 9.0,''
  Comput.\ Phys.\ Commun.\  {\bf 207} (2016) 432
%  doi:10.1016/j.cpc.2016.06.008
  [arXiv:1601.01167 [hep-ph]].
  %%CITATION = doi:10.1016/j.cpc.2016.06.008;%%
  %308 citations counted in INSPIRE as of 26 Mar 2020

\bibitem{Mathematica}
Wolfram Research, Inc., Mathematica, Version 12.1, Champaign, IL (2020), https://www.wolfram.com/mathematica

%\cite{Agostinelli:2002hh}
\bibitem{Agostinelli:2002hh}
  S.~Agostinelli {\it et al.} [GEANT4 Collaboration],
  %``GEANT4: A Simulation toolkit,''
  Nucl.\ Instrum.\ Meth.\ A {\bf 506} (2003) 250.
%  doi:10.1016/S0168-9002(03)01368-8
  %%CITATION = doi:10.1016/S0168-9002(03)01368-8;%%
  %10064 citations counted in INSPIRE as of 13 Feb 2019  


\bibitem{RenatGit}
https://gitlab.cern.ch/P348/na64-tools/tree/alps/geant4/simulation/examples/example100500


%\cite{Banerjee:2020fue}
\bibitem{Banerjee:2020fue}
  D.~Banerjee {\it et al.},
  %``Search for Axionlike and Scalar Particles with the NA64 Experiment,''
  Phys.\ Rev.\ Lett.\  {\bf 125} (2020) 8
   [Phys.\ Rev.\ Lett.\  {\bf 125} (2020) 081801]
 % doi:10.1103/PhysRevLett.125.081801
  [arXiv:2005.02710 [hep-ex]].
  %%CITATION = doi:10.1103/PhysRevLett.125.081801;%%
  %4 citations counted in INSPIRE as of 24 Aug 2020

  
\bibitem{charm}
F. Bergsma {\it et al.}, [CHARM Collaboration], Phys.\ Lett.\  B {\bf 157},  458 (1985).  
  
\bibitem{nucal}
J. Bl\"umlein {\it et al.}, 
%?Limits on neutral light scalar and pseudoscalar particles in a proton beam dump experiment,? 
Z. Phys. C {\bf 51}, 341 (1991).


%\cite{Aubert:2008as}
\bibitem{Aubert:2008as}
  B.~Aubert {\it et al.} [BaBar Collaboration],
  %``Search for Invisible Decays of a Light Scalar in Radiative Transitions $\upsilon_{3S} \to \gamma$ A0,''
  arXiv:0808.0017 [hep-ex].
  %%CITATION = ARXIV:0808.0017;%%
  %126 citations counted in INSPIRE as of 07 Apr 2020

\bibitem{e141b}
M.~W.~Krasny {\it et al.} [E141 Collaboration], 
%"Recent searches for pseudoscalar bosons in electron beam-dump experiments", 
Preprint Univ. of  Rochester, UR-1029 (1987); 

\bibitem{lep}
G. Abbiendi {\it et al.} [OPAL Collaboration], Eur. Phys. J. C {\bf 26}, 331 (2003).



\bibitem{primex}
D. Aloni, C. Fanelli, Y. Soreq, M. Williams, Phys.\ Rev.\  Lett.  {\bf 123},  071801 (2019).



%\cite{Dolan:2017osp}
\bibitem{Dolan:2017osp}
  M.~J.~Dolan, T.~Ferber, C.~Hearty, F.~Kahlhoefer and K.~Schmidt-Hoberg,
  %``Revised constraints and Belle II sensitivity for visible and invisible axion-like particles,''
  JHEP {\bf 1712} (2017) 094
 % doi:10.1007/JHEP12(2017)094
  [arXiv:1709.00009 [hep-ph]].
  %%CITATION = doi:10.1007/JHEP12(2017)094;%%
  %34 citations counted in INSPIRE as of 22 Jan 2019

%\cite{Dobrich:2019dxc}
\bibitem{Dobrich:2019dxc}
  B.~Dobrich, J.~Jaeckel and T.~Spadaro,
  %``Light in the beam dump. Axion-Like Particle production from decay photons in proton beam-dumps,''
  JHEP {\bf 1905} (2019) 213
%  doi:10.1007/JHEP05(2019)213
  [arXiv:1904.02091 [hep-ph]].
  %%CITATION = doi:10.1007/JHEP05(2019)213;%%
  %16 citations counted in INSPIRE as of 07 Apr 2020
  
%\cite{Knapen:2016moh}
\bibitem{Knapen:2016moh}
  S.~Knapen, T.~Lin, H.~K.~Lou and T.~Melia,
  %``Searching for Axionlike Particles with Ultraperipheral Heavy-Ion Collisions,''
  Phys.\ Rev.\ Lett.\  {\bf 118} (2017) no.17,  171801 
 % doi:10.1103/PhysRevLett.118.171801
  [arXiv:1607.06083 [hep-ph]].
  %%CITATION = doi:10.1103/PhysRevLett.118.171801;%%
  %79 citations counted in INSPIRE as of 14 Aug 2020  
  
%\cite{BelleII:2020fag}
\bibitem{BelleII:2020fag}
  [Belle-II Collaboration],
  %``Search for Axion-Like Particles produced in $e^+e^-$ collisions at Belle II,''
  arXiv:2007.13071 [hep-ex].
  %%CITATION = ARXIV:2007.13071;%%
  %1 citations counted in INSPIRE as of 14 Aug 2020
  
  
  
%\cite{Feng:2018noy}
\bibitem{Feng:2018noy}
  J.~L.~Feng, I.~Galon, F.~Kling and S.~Trojanowski,
  %``Axionlike particles at FASER: The LHC as a photon beam dump,''
  Phys.\ Rev.\ D {\bf 98} (2018) no.5,  055021
 % doi:10.1103/PhysRevD.98.055021
  [arXiv:1806.02348 [hep-ph]].
  %%CITATION = doi:10.1103/PhysRevD.98.055021;%%
  %38 citations counted in INSPIRE as of 14 Aug 2020
  

  

\end{thebibliography}
\end{document}